\documentclass[12pt]{article}
\usepackage{a4}
\usepackage[dvips]{graphicx}
\begin{document}

\centerline{\bf Notes about equivalence between the Sine -Gordon theory}
\centerline{ \bf (free fermion point) and the free fermion theory.}

\centerline{\bf S.V.Kryukov    }
\centerline{\it  Bogoliubov Laboratory of Theoretical Physics,}
\centerline{\it Joint Institute for Nuclear Research,}
\centerline{\it 141980, Dubna (Moscow region), Russia.}
\centerline{\it e-mail kryukov@itp.ac.ru,kryukov@thsun1.jinr.ru}
{\sf {\bf Abstract.}
The  space of local integrals of motion for the
Sine-Gordon theory (the free fermion point) and the theory of  free
fermions in the  light cone coordinates is investigated.
Some important differences
between the  spaces of  local integrals of motion of these theories are
obtaned.
 The equivalence is broken on the level of the integrals
of motion between bosonic and fermionic theories (in the free fermion
point). The integrals of motion are constructed without Quantum Inverse
Scattering Method (QISM) and the additional quantum
integrals of motion are obtained.So the  QISM is not
absolutely complete.}

\centerline{\bf Classification codes:PACS 02.10.-v,03.65.-w}

\centerline {\bf Keywords: quantum integrals
of motion,free fermions, Sine-Gordon }

\centerline{\bf 1. Introduction}

We have obtained equivalence between the Sine-Gordon theory
(the free fermion point) and free fermions by comparing the
expressions commuting with the  Hamiltonians of these theories in a space
of
integrated
expressions of local densities.  Equivalence of these
theories has been investigated by comparing the expressions  commuting
with the Hamiltonians
in a space of non integrated expressions of local densities.
We have obtained some important differences in this space.
In the subsequent paper  we will  investigate this equivalence
between arbitrary values of coupling constants.
In our opinion, we must present quantum integrals of
motion if we quantize the integrable models.
Quantum equivalence of the theories must be   checked also by
using quantum integrals of motion.
The procedure of quantization for an integrable theory,
which  has no  quantum integrals of motion, must not be cousistent [1,2].
We have obtained that the S-G theory (the free fermions point)
and the theory of  massive fermions are not equivalent  on the
level of the quantum integrals of motion.
So the question about massiveness of the S-G theory at this point is open.
Our construction of  quantum integrals of motion does not really use
the QISM.
Up to now in the literature (there are many  articles and even textbook
[5])
there is no  evidence of that QISM gives all  the integrals
of motion.

\centerline{\bf 2. Classical IM for free fermions}
Let us consider the classical model of  free fermions in  the light -cone  coordinates.
The equations of motion have the form
$$
i\partial_{-}\psi_{1}=m\psi_{2},~~~
i\partial_{+}\psi_{2}=m\psi_{1},\eqno(2.1)
$$
$$
-i\partial_{-}\psi_{1}^{+}=m\psi_{2}^{+},~~~
-i\partial_{+}\psi_{2}^{+}=m\psi_{1}^{+},
$$
where $\psi_{i},and \psi^{+}_{i}$  are the anti-commutative fields.
Now we see that  some equations (2.1) are  connections
(second type in Dirac terminology).
We choose the $x^{-}$  coordinate to represent the evolution time.
We will work in the Euclidean space.
We should like to  note that the functions,     which are contained in the
Hamiltonian
and  IM, are  initial functions
and do not depend on the time variable $x^{-}$. Obviously, they do not
satisfy any equation of motion.
The Poisson bracket is written for the  initial functions.
The operator of the time
evolution
(Hamiltonian)  has the form
$$
H=\frac{m}{2}\int_{-\infty}^{+\infty}dx(\psi_{1}(x)\psi_{2}^{+}(x)
+\psi_{2}(x)\psi_{1}^{+}(x)). \eqno (2.2)
$$
Now we must resolve the connection (2.1)
for the initial function $\psi_{2}(x)$.We will consider
$\psi_{1}(x)=\psi (x)$
as a dynamic variable and $\psi_{2}(x)$ as  a non dynamic variable.
The Poisson bracket between $\psi (x)$
and $\psi^{+} (x)$ has the form
$$
\{\psi(x),\psi^{+}(y)\}=\delta(x-y).  \eqno (2.3)
$$
Now  the Hamiltonian in these variables can be obtained as follows:
$$
H=\frac{m^{2}}{2}\int_{-\infty}^{+\infty}dx
[\psi (x)(\int_{-\infty}^{x}\psi ^{+}(t)dt-
\int_{x}^{+\infty}\psi^{+}(t))+
$$
$$
+\psi^{+}(x)(\int_{-\infty}^{x}\psi (t)dt-
\int_{x}^{+\infty}\psi (t)dt)], \eqno (2.4)
$$
 It is a simple  exercise to check that the classical IM has the form
$$
I_{n}=\int_{-\infty}^{+\infty}dx(\psi(x)\partial_{x}^{n}\psi^{+}(x)+h.c.),~~~n=1,2,3...
\eqno(2.5)
$$
We use the Poisson bracket (2.3) to check the following Poisson brackets:
$$
\{ H,I_{n} \}=0,~~~\{I_{n},I_{m}\}=0. \eqno(2.6)
$$
\centerline{\bf 3.Quantum IM for free fermions}

The first step in the  quantization of the IM is  the following. We note that the functions
$\psi_{i}$ and $\psi_{i}^{+}$ (which are present in the  IM and in the $H$)
do not really  depend on the  evaluation time  because they represent
the initial data.
In the light - cone coordinates this  is equivalent to the holomorphic condition
(in the 2D Euclidean space).The
$x^{-}$- coordinate is determined by  $H$, and cannot be defined before
 $H$  is  determined.
So we have
$$
\partial_{-}\psi_{i}(x^{+},x^{-})=0,~~~i=1,2,
$$
and the same condition for   $\psi^{+}$.
It is absolutely wrong to consider this condition as an  equation of
motion.
The second step is to note that we can represent the
initial data as  Laurent series,
$$
\psi_{i}(x)=\sum_{n}\psi_{in}x^{-n},~~~
\psi_{i}^{+}(x)=\sum_{n}\psi^{+}_{in}x^{-n-1},
$$
and define  $H$ and   IM like as  ordering operators.
 The third step is to define the quantum commutators in our
system. Let us  obtain the connection for Laurent modes $\psi_{n}$.
We introduce the following
integrals:
$$
\psi_{n}(l)=\int_{-\infty}^{+\infty}dx\psi(x)  (x-l)^{n-1},~~~
\psi_{n}^{+}(l)=\int_{-\infty}^{+\infty}dx\psi^{+}(x) (x-l)^{n},
$$
($\psi_{1}(x)=\psi(x)$ for simplicity), where $l$ is an arbitrary coordinate.
The contour of  integration  can be chosen as  in  Fig. 1.
\begin{figure}[h]
\centering\includegraphics[height=4cm]{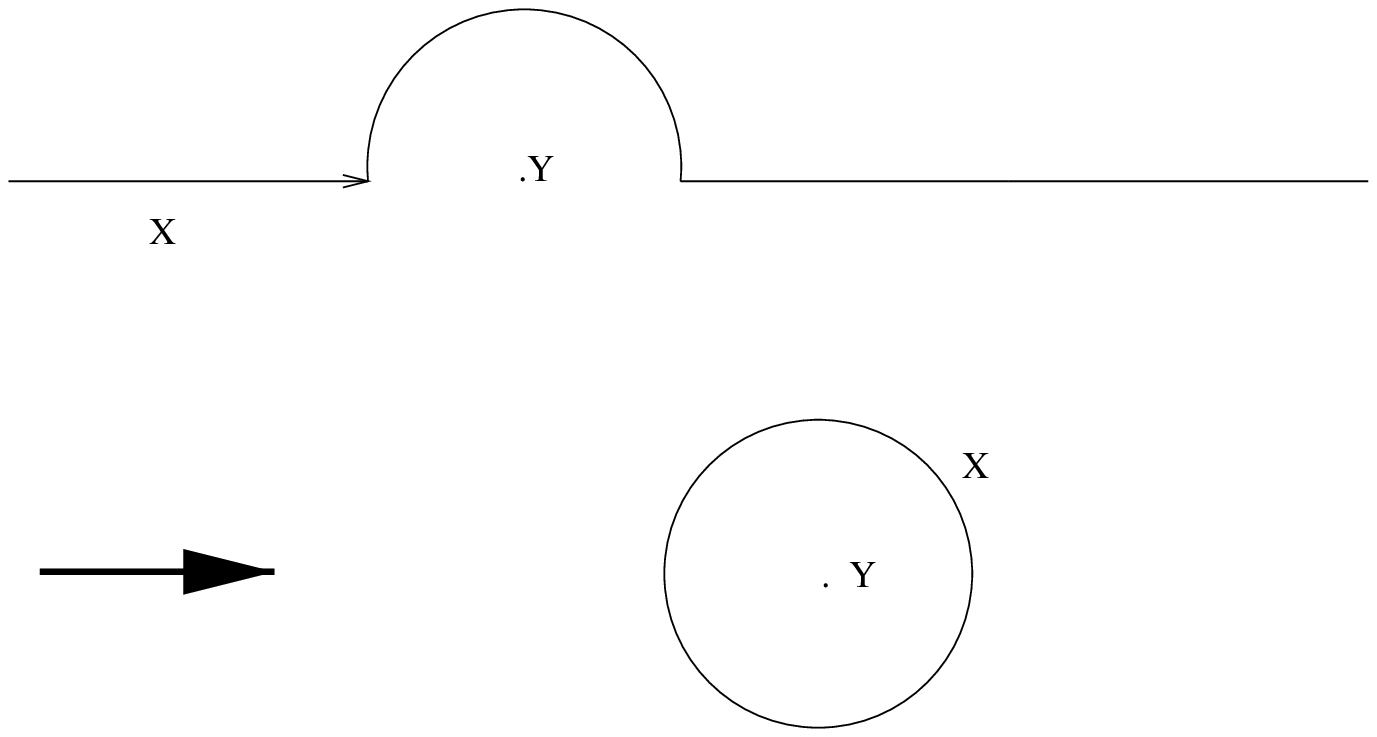}
\centerline{\small { Fig.1 Contour of the  integration for the expandsion.}   }
\end{figure}
We must deform  the integration contour like in  Fig.1 and obtain the
following:
$$
\psi_{n}(l)=\oint_{l} dx \psi(x)(x-l)^{n}, \eqno (3.1)
$$
i.e., the ordinary coefficients of the  Lourent expansion. Now we have
$$
\{\psi_{n}(l),\psi_{m}^{+}(l)\}=\int_{-\infty}^{+\infty}dx
(x-l)^{n-1}\int_{-\infty}^{+\infty}dt(t-l)^{m}\delta(x-t)=
$$
$$
=\int_{-\infty}^{+\infty}dx(x-l)^{n+m-1}=
\oint_{l} dx \frac{1}{(x-l)^{-n-m+1}}=\delta_{n+m,0}.
$$
We can always take $l=0$ and obtain
$$
\{\psi_{n},\psi^{+}_{m}\}=\delta_{n+m,0}.
$$
Now we will quantize the bracket replacing the Poisson bracket with the anticommutator
 of the fields.
Let us  introduce   ordering of the operators  $\psi_{n}$
as  follows:
$$
\psi_{n}\psi_{-n}^{+}=:\psi_{n}\psi^{+}_{-n}:+1,~~~n>0~~~
\psi_{-n}^{+}\psi_{n}= :\psi_{-n}^{+}\psi_{n}:.
$$
$$
\psi_{-n}\psi_{n}^{+}=:\psi^{+}_{n}\psi_{-n}:,~~~
\psi_{n}^{+}\psi_{-n}= :\psi_{-n}\psi_{n}^{+}:+1
$$
It is a very simple exercise to check that
$$
\psi(x)\psi^{+}(y)=\frac{1}{(x-y)} +:\psi(x)\psi^{+}(y):,~~~~
\psi^{+}(x)\psi(y)=\frac{1}{(x-y)} +:\psi^{+}(x)\psi(y):.
$$
Of course,  there are many ways of ordering these operators,
but if we want to preserve quantum integrability we must choose the  one above.
Probably there are other possibilities.

Now  let us  determine the quantum Hamiltonian and  the quantum IM.
It is not a problem to  write the quantum Hamiltonian
$$
H=\frac{m^{2}}{2}\int_{-\infty}^{+\infty}
dx(:\psi(x)(\int_{-\infty}^{x}\psi^{+}(t):-\int_{x}^{+\infty}\psi^{+}(t):)dt+
:\psi(x)^{+}(\int_{-\infty}^{x}\psi(t):-\int_{x}^{\infty}\psi(t):)dt\eqno(3.2)
$$
and
$$
I_{n}=\int_{-\infty}^{+\infty}P_{n}[\psi(x),\psi^{+}(x)]dx,
$$
where $P_{n}$ are the ordering differential polynomials
of  $\psi(x)$ and $\psi^{+}(x)$.
The coefficients of these polynomials are determined by  the conditions
$$
[H,I_{n}]=0.\eqno(3.3)
$$
The quantum field $\psi(x^{+},x^{-})$
(solution of the quantum equation of motion) has the form
$$
\psi(x^{+},x^{-})=\exp iHx^{-}\psi(x^{+})\exp-iHx^{-},
$$
and this representation leads to the ordinary equal time anticommutator.
Now we must determine the way of calculating  the commutator.
Let us to  choose the contour of integration in (3.4) as in Fig.2.
$$
[\int_{-\infty}^{+\infty} h(x)dx;i_{n}(y)]=
\int_{-\infty}^{+\infty} h(x) i_{n}(y)dx
-\int_{-\infty}^{+\infty}i_{n}(y) h(x)dx,\eqno(3.4)
$$
where $h(x)$  and $i_{n}(x)$  are the densities of  $H$ and $I_{n}$.

\begin{figure}[h]
\centering\includegraphics[height=1.5cm]{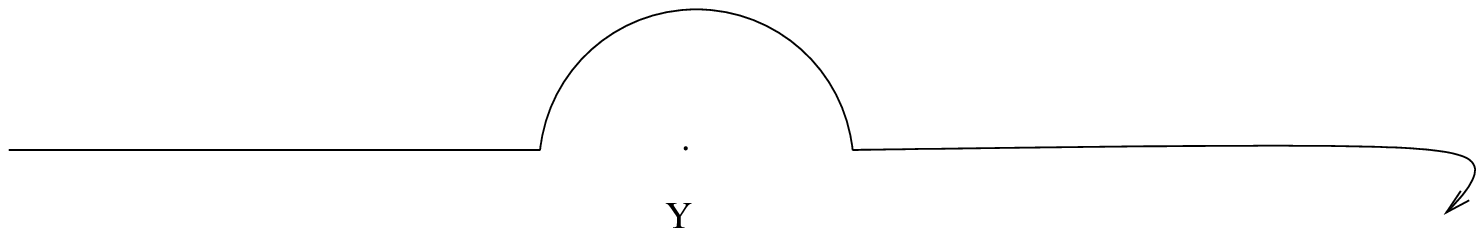}
\centerline {\small Fig.2a Contour of   integration for the commutator}
\end{figure}
The contour of  integration of the fist integral (3.4) is shown in   Fig.
2a
and the contour of  integration of the second integral (3.4) is shown in
Fig
2b. Now we can add $C_{R}$,
or  continue  the integration contour analytically.
\begin{figure}[h]
\centering\includegraphics[height=1cm]   {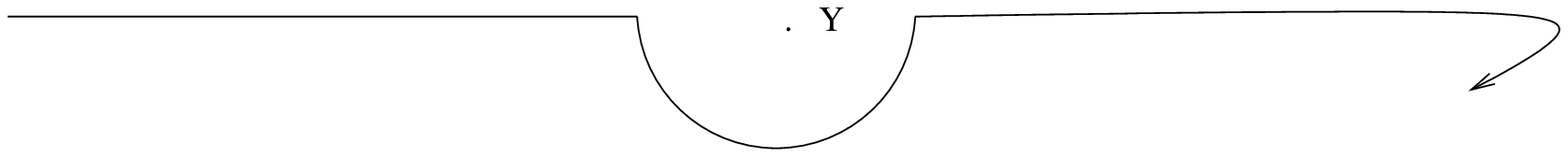}
\centerline {\small  Fig.2b Contour of   integration for the commutator }
\end{figure}
Therefore,  we have
$$
[H,i_{n}(y)]=\oint_{y}dxh(x)i_{n}(y).\eqno(3.5)
$$
Of course, there is another singular point
$l$ which is  the center of expansion for the
field  $\psi(x)$ (see (3.1)) which does  not have any physical meaning.
After implementing
the procedure (3.5), we must obtain the  full difference. Now we represent
IM in  the form
$$
I_{n}=\int_{-\infty}^{+\infty}
dx (:\psi(x)\partial^{n} \psi^{+}(x):+h.c.)~~~n=1,2,3....\eqno(3.6)
$$
They (IM) are very similar to the  classical ones, because we use only one
contraction ( Poisson bracket) for calculating $I_{n}$.
For Massive Thirring model we will have more complicated contractions, and
$I_{n}$
will be different from classical ones. Now we can obtain  the involution
$$
[I_{n},I_{m}]=0,
$$
where the rule of calculation of the commutator has been  introduced
above.

\centerline{\bf  4. Quantum IM for Sine-Gordon equation}

Quantum IM for the Sine -Gordon theory can be obtained  in a similar
way.
We have holomorphy condition for the initial function
$$
\partial_{-}\varphi(x^{+},x^{-})=0.
$$
The holomorphic function can be expanded as ($x^{+}=x$)
$$
\varphi(x)=q+p\log x- \sum_{n \ne 0}\frac {a_{n}x^{-n}}{n}.
$$
We consider the boundary condition  $\partial \varphi (x) \to 0$
for $x\to \pm \infty$.The
Poisson bracket and the  Hamiltonian have the form
$$
\{\varphi(x);\varphi(y)\}=\pi \varepsilon (x-y),~~~
H=\int _{-\infty}^{+\infty}\exp \alpha \varphi(x)+
\exp -\alpha \varphi (x)dx.
$$
Let us introduce the following commutation relations
$$
[\hat{p},\hat{q}]=1,~~~ [\hat{a}_{n},\hat{a}_{m}]=n\delta_{n+m,0}.
$$
The  ordering of these operators can be determined by
$$
\hat  p \hat  q=:\hat  p \hat q:+1,~~~
\hat  q \hat  p=:\hat  p \hat q:,
$$
$$
\hat  a_{n}\hat  a_{-n}=:\hat  a_{n} \hat  a_{-n}:+n,~~~
\hat  a_{-n}\hat  a_{n}=:\hat  a_{-n} \hat  a_{n}:,~~~n>0.
$$
So we have
$$
\varphi(x)\varphi(y)=:\varphi(x)\varphi(y):+\log(x-y).\eqno(4.1)
$$
From (4.1) we can obtain
$$
[\varphi(x);\varphi(y)]=i\pi\varepsilon(x-y).
$$
So we have quantized the above Poisson bracket,  and the
quantum Hamiltonian has the form
$$
H=\int_{-\infty}^{+\infty}
:\exp \alpha\varphi(x):+:\exp -\alpha\varphi(x):dx.
$$
The rules of calculation for the commutators are the same as for the
 fermion theory.
Now we can obtain the simplest IM for an  arbitrary position of the coupling constant
$$
i_{2}(x)=:(\partial\varphi(x))^2:,
$$
$$
i_{4}(x)=:(\partial\varphi(x))^4:+
\frac{(\alpha^{4}-6\alpha^{2}+4)}{\alpha^{2}}:(\partial^{2}\varphi(x))^2:,
\eqno(4.2)
$$
$$
i_{6}(x)=:(\partial\varphi(x))^6:+
\frac{5(-\alpha^{4}+8\alpha^{2}-4)}{3\alpha^{2}}
     :(\partial\varphi(x))^3\partial^{3} \varphi(x):+
$$
$$
+\frac{(3\alpha^{8}-40\alpha^{6}+
155\alpha^{4}-160\alpha^{2}+48)}{6\alpha^{4}}
:(\partial^{3}\varphi(x))^2:.
$$
We can check the classical limit ($\alpha \to 0$) [4] and  the
commutativity of these
operators
$I_{n}$ with the  densities (4.2).
For  special position of the coupling constant $\alpha=1$
(the free fermion point) we have more integrals of motion
$$
i_{2 }(x) = :(\partial \varphi (x)  ) ^{2 }:,
$$
$$
i_{3 }(x) = :(\partial \varphi (x)  )^{3 }:, \eqno(4.3)
$$
$$
i_{4 }(x) = : (\partial \varphi (x)  )^{4 }:-6:(\partial \varphi (x)  )^{2 }
\partial ^{2 } \varphi (x) : +
4:\partial \varphi (x) \partial ^{3 } \varphi (x) :+
$$
$$
+3:(\partial ^{2} \varphi (x)^{2}  ): - \partial ^{4 } \varphi (x),
$$
$$
i_{5 }(x) = : (\partial \varphi (x)  )^{5 }:
- 10 : (\partial \varphi (x)  ) ^{3 }\partial ^{2 } \varphi (x) : +
$$
$$
+ 10 : (\partial \varphi (x)  ) ^{2 }
\partial ^{3 }\varphi (x) :
- 10 : \partial ^{2 } \varphi (x)  \partial ^{3 } \varphi (x) : +
$$
$$
+\partial ^{5 } \varphi (x)+
5 : (3(\partial ^{2 } \varphi (x) ) ^{2 } -
\partial ^{4 } \varphi (x) )  \partial \varphi (x):.
$$
We can check the commutativity of the $I_{n}$   operators with the densities (4.3)
 [7]. Now we see some connection
 between IM
(3.6)   and   (4.3). Let us introduce some equivalence between $\psi (x)$
and $:\exp\varphi(x):$
($\psi(x)\sim:\exp\varphi(x):$  and $\psi^{+}(x)\sim:\exp-\varphi(x):$)
where  $\psi(x)$, $\psi^{+}(x)$ are the
initial functions in the  fermion theory and $\varphi(x) $ is the
initial function in the Sine-Gordon theory. After some algebra we see that
IM
(3.6) and (4.3) are the same. So we must consider that   $\alpha=1$
is the free fermion point in the  Sine-Gordon theory.
We see that $i_{3}$ and $i_{5}$ are new integrals of motion, which cannot
be
obtained by QISM, so we can conclude that the QISM is not a complete
method.
Another interesting note consists in the following.
Where the completeness of the QISM is broken,we have a new
infinite dimensional symmetry that is very similar to  some reduction
of the
$W_{1+\infty}$ algebras obtained in [6].
If we have $\alpha =1$, we can obtain local conservation currents
in the  Sine-Gordon theory
$$
[H,J^{i}(x)]=0,
$$
where
$$
J^{i}(x)=\frac{1}{(i+1)}:(\partial^{i+1}\exp\varphi(x))\partial \exp
-\varphi(x):+
\frac{1}{(i+1)(i+2)}:(\partial^{i+2}\exp \varphi(x))\exp-\varphi(x):,
$$
where $i\in N$.We have  holomorphic operators $J^{i}(x).$
Now we can calculate the commutators of  Laurent  modes of these
currents and obtain
$$
[J_{n}^{k},J_{m}^{p}]=
(k+1)!\sum_{l'=0}^{k} \frac{  J_{n+m}^{p+l'}}{l'!(k+1-l')!}
\prod_{j'=0}^{ k-l'}(m+p+1-j')-
$$
$$
-(p+1)!\sum_{l=0}^{p} \frac{  J_{n+m}^{k+l}}{l!(p+1-l)!}
\prod_{j=0}^{p-l}(n+k+1-j)
$$
$$
$$
$$
+\delta_{n+m,0}\frac{(p+1)!(k+1)!}{2(k+p+3)!}\times
$$
$$
\times ((-1)^{k+1}\prod_{j=0}^{2+k+p}(n+k+1-j)- (-1)^{p+1 }
\prod_{j=0}^{2+k+p}(m+p+1-j) ),
$$
where we have
$J^{i}_{n} = \frac{1}{2\pi i}\oint_{0} J^{i}(x)x^{i+1+n}dx$.

The terms in $H$ for free fermions prohibit the existence of the currents
(4.4)
in the fermion theory. So we see the absence of  equivalence between the
massive
fermion theory and the free fermion point for the Sine-Gordon theory.

\centerline{\bf 5. Conclusion}

If we do not obtain equivalence between the theory of free fermions
and the Sine- Gordon theory (in free fermions point), so we can not
consider
Sine- Gordon theory as  a massive theory.
The existence of  local conservation currents also confirms this fact.
It is very interesting to solve this model in the  "free fermions" point
using the new infinite dimensional symmetry which is very similar to
conformal
symmetry.

Another important note of the article consists in the following.
We have  constructed the example where the QISM does not  give  all
the
integrals of motion. There are many articles  about QISM, but the
problem of  completeness of this method has not been discussed,
so our example  appears very important.
\centerline{\bf  References}

[1] Coleman ~S 1975  {\it Phys Rev D} {\bf 11} 2088

[2] Mandelstam ~S 1975 {\it Phys Rev D} {\bf 11} 3026

[3] Kryukov ~S 1996 {\it JETP Lett} {\bf 63} 375

[4] Takhtadzan ~L 1974 {\it JETP} {\bf 66} 474

[5] Faddeev ~L,~Sklyanin ~E, ~Takhtajan ~L 1980
{\it Theor Math Phys} {\bf40}
688;

Takhtadzan ~L, ~Faddeev ~L, Hamilton aproach in soliton theory Moscow 1986

[6] Bakas ~I, ~Khesin ~B, ~Kiritsis 1993 {\it Comm Math Phys} {\bf 151} 233

\end{document}